\newif\iftwocol
\twocolfalse

\iftwocol
    \documentclass[10pt]{iopart} 
\else
    \documentclass[12pt]{iopart} 
\fi

\usepackage{xr}
\expandafter\let\csname equation*\endcsname\relax
\expandafter\let\csname endequation*\endcsname\relax
\usepackage{amsmath}

\usepackage{acronym}
\usepackage{graphicx}
\usepackage{chemformula}
\usepackage{multirow}
\usepackage{tabularray}
\usepackage[style=numeric-comp,sorting=none]{biblatex} 
\usepackage{hyperref} 

\newcommand{\shortTitle}{Ni-CN on 3D Py-C Electrodes for Enhanced Hydrogen Evolution}
\newcommand{\longTitle}{Conformally-Coated Nickel-Carbon Nitride on 3D Structured Pyrolytic Carbon Electrodes for Enhanced Hydrogen Evolution}
\newcommand{\CustomDate}{June 2025}

\newcommand{\ECSAoverGSAacidicCN}{85}
\newcommand{\ECSAoverGSAacidicNi}{10}
\newcommand{\ECSAoverGSAacidicBare}{1}
\newcommand{\ECSAoverGSAalkalineCN}{77.5}
\newcommand{\ECSAoverGSAalkalineNi}{24}
\newcommand{\ECSAoverGSAalkalineBare}{2.5}

\newcommand{\mAcmsquared}{mA\textperiodcentered{}cm\textsuperscript{-2}}

\newacro{HER}{hydrogen evolution reaction} 
\newacro{RCD}{rod-connected diamond} 
\newacro{PyC}{pyrolytic carbon} 
\newacro{ECSA}{electrochemical active surface area} 
\newacro{CN}{carbon nitride} 
\newacro{GSA}{geometric surface area} 
\newacro{CV}{cyclic voltammetry} 
\newacro{3D-PyC}{3D pyrolytic carbon} 

\DeclareUnicodeCharacter{2212}{-}

\newcommand*\sfref[2][]{figure~\ref*{#2}#1 in the supplementary materials}
\newcommand*\sFref[2][]{Figure~\ref*{#2}#1 in the supplementary materials}
\newcommand*\stref[2][]{table~\ref*{#2}#1 in the supplementary materials}
\newcommand*\sTref[2][]{Table~\ref*{#2}#1 in the supplementary materials}

\begin{document}

\title[\shortTitle{}]{\longTitle{}}

\author{
Nadira Meethale Palakkool$^{1}$,
Mike P. C. Taverne$^{1,2,\ast}$,
Owen G. Bell$^{1}$,
Christopher P. Jones$^{3}$,
Jonathan D. Mar$^{4}$,
Duc Tam Ho$^{1}$,
Vincent Barrioz$^{1}$,
Yongtao Qu$^{1}$,
Zhong Ren$^{5}$,
Chung-Che Huang$^{6,\ast}$,
Ying-Lung Daniel Ho$^{1,2,\ast}$
}

\address{$^{1}$ School of Engineering, Physics and Mathematics, Northumbria University, UK}
\address{$^{2}$ School of Electrical, Electronic and Mechanical Engineering, University of Bristol, Bristol, UK}
\address{$^{3}$ Interface Analysis Centre, H.H. Wills Physics Laboratory, University of Bristol, Bristol, UK}
\address{$^{4}$ School of Mathematics, Statistics and Physics, Newcastle University, Newcastle upon Tyne, UK}
\address{$^{5}$ Oxford Instruments Plasma Technology, Govier Way, Severn Beach, Bristol, UK}
\address{$^{6}$ School of Electronics and Computer Science, University of Southampton, Southampton, UK}

\ead{
mike.taverne@northumbria.ac.uk;
cch@soton.ac.uk;
daniel.ho@northumbria.ac.uk
}

\vspace{10pt}
\begin{indented}
\item[]\CustomDate
\end{indented}

\begin{abstract}
This study presents a three-dimensional electrode for the hydrogen evolution reaction (HER), tackling challenges in corrosion resistance, catalytic activity, and durability. The electrode features a rod-connected diamond structure made of pyrolytic carbon (PyC), with sequential Ni and CN conformal coatings, resulting in an increase of its electrochemical active surface area more than 80 times its geometric surface area, marking the first application of CN/Ni/RCD-PyC electrodes for HER. The CN layer enhances corrosion resistance under acidic and alkaline conditions, while synergizing with Ni boosts catalytic activity. We demonstrate that the CN/Ni/RCD-PyC electrode exhibited superior HER activity, achieving lower overpotentials of 0.6 V and 0.4 V for a current density of 10~mA\textperiodcentered{}cm\textsuperscript{-2} in acidic and alkaline media, respectively. The electrode demonstrates excellent durability with performance over 2000 cycles of cyclic voltammetry at high scan rates with no significant decay. These innovations result in significantly improved catalytic performance, including superior cyclic stability, low overpotential, and high-current density. This study provides a fresh perspective on 3D electrode development, offering valuable insights into material design for efficient and robust HER electrocatalysts and contributing to the advancement of sustainable hydrogen production technologies.

\end{abstract}

%
%
%
%
\iftwocol
    \ioptwocol
\fi

\section{Introduction}

As global energy consumption rises, hydrogen is increasingly seen as a key solution due to its high energy density, low carbon emission, and excellent convertibility \cite{Zhang2024ATechnologies}.
Electrocatalytic water splitting is a promising method for the production of green hydrogen, but its high cost is often driven by the use of noble metals as electrocatalysts\cite{Li2020MetallicSplitting,Gong2022PerspectiveSplitting,Zhao2023Chemical-vapor-deposition-grownEvolution}.
In this context, 3D-printing technologies have become a viable method for fabricating electrodes for electrochemical applications\cite{Zou2024AdditiveH2O2}. Advances in hierarchical electrode structure using 3D-printing, demonstrate that optimizing surface area-to-volume ratio (SA:V) significantly enhances mass transport and electrochemical performance\cite{Zhu2018TowardPrinting}.
Due to their excellent versatility and abundance, carbon materials are increasingly replacing precious metals.
3D-printed carbon materials, with their precise processability, facilitate the creation of intricate designs that are difficult to achieve using conventional manufacturing techniques\cite{Huner2022ElectrodepositionMedia,Peng20203DSplitting}.
The fusion of 3D-printing with traditional methods has enabled precise customization of porous carbon macro- and microstructures, significantly enhancing efficiency and optimization for specific applications\cite{Romero2024ATechniques}.
However, traditional 3D-printing methods face challenges, including layer-to-layer interface weaknesses, the need for extensive post-processing, limited resolution, and constraints with high-temperature materials\cite{Oztan2021UtilizationReview}.

Recently, converting 3D-printed polymers into carbon-based material through pyrolysis has gained attention\cite{Pan2022HybridManufacturing,Zou2024AdditiveH2O2}.
This method enhances the electrode performance and overcomes the limitations of conventional printing.
For instance, Rezaei et al. demonstrated a \acf{3D-PyC} electrode produced by stereolithography (SLA) and pyrolysis, offering promising results for electrochemical applications\cite{Rezaei2020HighlyTechnology}.
Amorphous \ac{PyC} derived from polymeric precursors offers several advantages for electrochemical applications, including a wide potential stability window, low cost and high electrocatalytic activity, making them a promising material for \acf{HER} applications\cite{Quang2019ElectrochemicalPolymers,McCreery2008AdvancedElectrochemistry}. 
Inspired by this, a similar method has been applied for the 3D-printing of polymer followed by pyrolysis under vacuum conditions.
Furthermore, this study demonstrates a novel electrode design for the 3D electrode using a diamond-crystal-inspired geometry based on an \acf{RCD} structure\cite{Taverne2022StronglyCrystals,Chen2019ObservationChalcogenides,Taverne2018,Chen2017DirectSpectroscopy,Taverne2016ModellingCrystals,Chen2015EvidenceStructures}. 

Recent reviews highlight the transformative potential of integrating 3D-printing with surface functionalization techniques to enhance the active surface area of the electrode.
These approaches leverage non-noble metal electrocatalysts and emerging two-dimensional (2D) materials such as transition metal carbides/nitrides (MXenes) and transition metal dichalcogenides (TMDs), offering significant advancements for HER applications\cite{MeethalePalakkool2025RecentSplitting}.
According to the Sabatier principle, an ideal electrocatalyst exhibits a Gibbs free energy of hydrogen adsorption ($\Delta \textnormal G_{\textnormal{H*}}$) close to zero, balancing the hydrogen binding strength\cite{PerezBakovic2021ElectrochemicallyElectrolyte}.
Nickel (Ni) is a promising alternative to noble metals for HER, with a $\Delta \textnormal G_{\textnormal{H*}}$ value close to platinum (Pt)\cite{Fu2014Acid-resistantNickel,Fu2018CarbonAdsorption}.
However, pure Ni is prone to dissolution and poisoning in acidic or alkaline environments\cite{Putri2023EngineeringPerspective,Wang2015Molybdenum-Carbide-ModifiedReaction}.  
Strategies such as encapsulating Ni in carbon cages or graphene layers have improved stability\cite{Deng2015EnhancedReaction,Tavakkoli2015SingleShellReaction}.
Furthermore, \acf{CN} offers high chemical and thermal stability, but suffers from poor conductivity and weak water dissociation in alkaline media\cite{Kumar2023MetalElectrocatalysis,Jin2018ConstructingEvolution,Kays2024AElectrolysis}. 
Encapsulating Ni in CN enhances stability and catalytic performance, making it a promising candidate for HER, especially when structure and surface geometry are carefully controlled\cite{Fu2014Acid-resistantNickel}. 

This study investigates a novel approach to enhancing the electrocatalytic performance of pyrolyzed 3D-printed carbon electrodes for HER.
The electrodes were fabricated using SLA 3D-printing, followed by vacuum pyrolysis to achieve a 3D-PyC structure, based on RCD, known as RCD-PyC.
Surface functionalization was accomplished through a sequential conformal deposition technique, involving galvanostatic electrodeposition to deposit Ni nanoparticles, followed by chemical vapor deposition (CVD) for a CN coating. 
The integration of Ni and CN exploits strong interfacial interactions and a synergistic catalytic mechanism, enhancing hydrogen adsorption and water dissociation in acidic and alkaline media\cite{Zhang2025ManipulationReaction,Chen2024MechanismReaction,R.S.2017MetalElectrocatalysts,Gao2022AAnode}.
The CN layer provides abundant nitrogen sites, improving the intrinsic activity and stability of Ni. To the best of our knowledge, this is the first application of 3D-PyC electrodes functionalized with Ni and CN for HER electrocatalysis.
The electrodes were characterized using scanning electron microscopy (SEM), energy-dispersive X-ray spectroscopy (EDS), X-ray diffraction (XRD), X-ray photoelectron spectroscopy (XPS) and Raman spectroscopy. Electrochemical testing demonstrated their robust activity and stability across different electrolyte conditions.
This work highlights the potential of combining 3D-printed carbon structures with advanced surface functionalization for scalable HER applications.

\section{Results and Discussion}

\subsection{3D RCD structured electrode design}\label{sec:geometry}

\begin{figure}
    \centering
    \includegraphics[width=\linewidth]{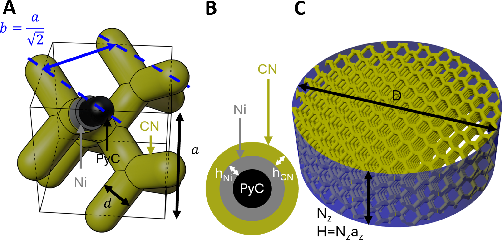}
    \caption{\textbf{Design of the 3D electrode:} (A) Unit-cell of the geometry used.
    It consists of 16 cylindrical rods positioned in a cube of size $a$ (the lattice constant).
    The diameter of the rods is $d$.
    The distance between lines of rods when viewed from the top is $b=a/\sqrt{2}$.
    (B) Cross-section through one of the rods after coating the \emph{RCD-PyC} template with \emph{Ni} and then \emph{CN}.
    The coating thicknesses of \emph{Ni} and \emph{CN} are $h_{Ni}$ and $h_{CN}$ respectively.
    (C) A complete crystal consisting of multiple unit-cells, truncated to a cylinder of diameter $D$. The height of the crystal is $H$ with a corresponding number of periods along the vertical direction $ N_z = H / a_{z}$.
    This template corresponds to the one submitted for 3D printing and has $N_z=3$ layers.}
    \label{fig:geometry}
\end{figure}

As illustrated in Figure \ref{fig:geometry} (A and B), the electrodes consist of an internal scaffold of PyC, coated with a layer of Ni and then a layer of CN.
The electrode geometry was based on the \emph{\acf{RCD}} geometry\cite{Taverne2022StronglyCrystals,Chen2019ObservationChalcogenides,Taverne2018,Chen2017DirectSpectroscopy,Taverne2016ModellingCrystals,Chen2015EvidenceStructures}.
As the name indicates, it is based on the atomic lattice of a diamond, where carbon atoms are arranged in a face-centerd cubic (FCC) lattice.
The RCD geometry is then created from this lattice by creating cylindrical rods, corresponding to the covalent bonds between the atoms. 
The specific 3D electrode employed in this work will henceforth be termed RCD-PyC throughout the text, and electrodes sequentially modified with Ni and CN as Ni/RCD-PyC and CN/Ni/RCD-PyC, respectively.

Figure \ref{fig:geometry}A shows a rectangular cuboid unit-cell of the resulting geometry.
It consists of 16 cylindrical rods, which can be divided into four identical tetrahedral arrangements of 4 rods each.
The electrodes are then generated by creating a $10\times10\times3$ array of this unit cell, which is then truncated to a circular shape.
Figure \ref{fig:geometry}C shows the resulting complete 3D electrode geometry of this process.
\Acf{PyC} lattice structures exhibit excellent mechanical properties, including the highest specific strength up to  4.42 GPa g${^{-1}} $cm${^3}$\cite{Bauer2016ApproachingCarbonnanolattices} and ultrahigh stiffness up to 21.6 GPa${^2}$\cite{Crook2020Plate-nanolatticesStrength}, due to the combination of the intrinsic properties of pyrolytic carbon and the designed lattice. The stretching-dominated lattice structures result in high strength and stiffness\cite{AshbyM.F2006TheLattices}, while bending-dominated structures provide high flexibility and mechanical energy absorptions\cite{Wagner2019ProgrammableTopologies}. Our RCD lattice exhibits both stretching and bending modes\cite{Chen2021CompressionRelationship}, allowing the optimization of strength, stiffness, flexibility, and energy absorption, which is essential for the electrodes.
Additionally, Ni coating can fill or eliminate surface cracks in the pyrolytic carbon lattice structure, enhancing its structural integrity and improving fracture toughness. The CN coating further strengthens the structure and improves wear resistance. The use of coatings to significantly enhance mechanical properties has been previously reported\cite{Bauer2014High-strengthMicroarchitecture}. 

Typically, the current density is plotted against the potential to account for the \acf{GSA} of the catalyst\cite{Anantharaj2018PrecisionAssessment}.
However, the surface area of the catalyst used in this study could not be reliably measured using the Brunauer-Emmer-Teller (BET) method, as the electrocatalyst surface area falls outside the lower limit of BET measurement, leading to poor gas adsorption.
However, the periodic geometry of the 3D-printed electrodes allows surface area calculation through SEM measurement and the creation of analogous CAD files.

\begin{figure}[b]
    \centering
    \includegraphics[width=\linewidth]{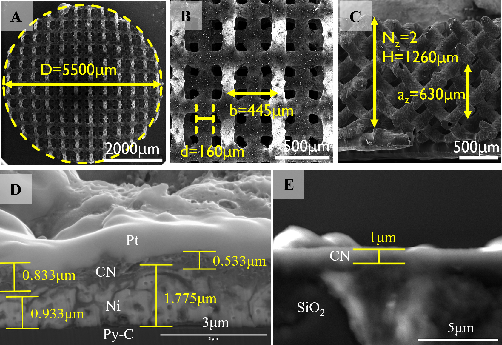}
    \caption{
    \textbf{SEM images of the  RCD-PyC electrode, with annotated measurements:}
    (A-D) The  RCD-PyC electrode (after coating with Ni and CN): (A) Overview, (B) Higher magnification top and (C) side view respectively.
    (D) SEM image of a cross-section, made using a Focused Ion Beam (FIB).
    See Fig. \ref{fig:geometry} for a description of the parameters and geometry.
    The thickness of the Ni layer is $\sim0.9-1.8\mu{}m$, while the thickness of the CN layer is $\sim0.5-0.8\mu{}m$.
    (E) Cross-section of an \ch{SiO2} on Si substrate that was coated with CN at the same time as the 3D electrode sample shown in (D).
    It shows a CN layer thickness of $\sim1\mu{}m$.
    }
    \label{fig:SEM+EDX}
\end{figure}

Hence, after fabrication, the dimensions of the resulting structure were estimated using SEM images (Figure \ref{fig:SEM+EDX}).
Using these dimensions, CAD models were created, from which the surface area could then be estimated.
\sTref{supp-tab:S1_geometry-parameters} shows the parameters used for the CAD models and the corresponding total surface areas of the electrodes: the original template ($GSA=20~cm^2$) and the template immediately after annealing ($GSA=2~cm^2$).
The change in \ac{GSA} when assuming a uniform flat coating with Ni and CN is negligible, due to the thin coating thicknesses (Ni thickness: 0.9-1.8\textmu{}m, \ac{CN} thickness: 0.5-0.8\textmu{}m), so the same $GSA=2~cm^2$ was used for calculations related to the bare and coated electrodes.


\subsection{Material Characterization}
\begin{figure}[b]
    \centering
    \includegraphics[width=\linewidth]{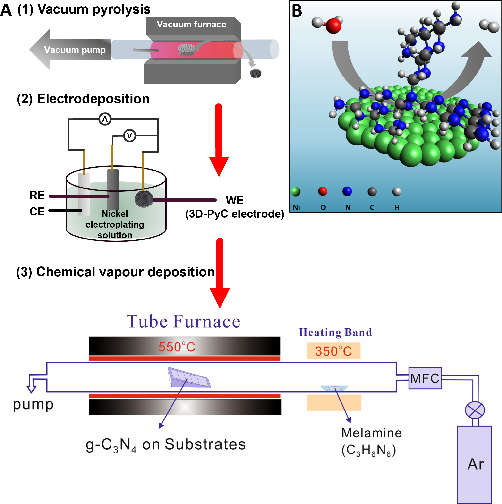}
    \caption{\textbf{Schematic representation of the fabrication process and HER mechanism.} (A) Illustration of the fabrication process including setups for vacuum pyrolysis, Ni electrodeposition and conformally CVD-grown CN. (B) Proposed HER mechanism involving hydrogen adsorption facilitated by spillover/migration effect via Ni and CN.}
    \label{fig:3}
\end{figure}
Demonstrating the geometric freedom of 3D-printing technologies and the high convertibility of polymeric materials simultaneously, we fabricated 3D-PyC electrodes with well-defined RCD geometry.
First, 3D-printing of the free-standing polymer structure was conducted via SLA, using high-temperature V2 resin.
Figure \ref{fig:3}A illustrates the schematic diagram of the fabrication process, including the surface functionalization.

\begin{figure}[b]
    \centering
    \includegraphics[width=\linewidth]{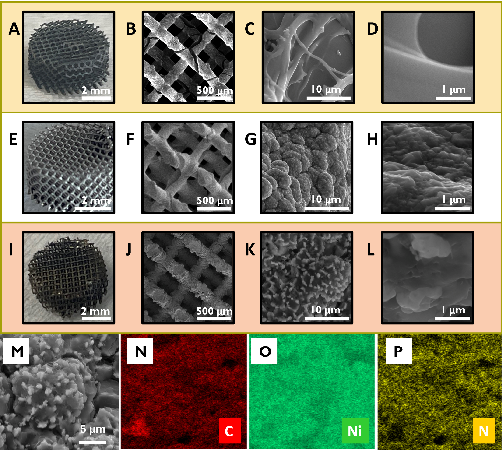}
    \caption{\textbf{Optical and SEM images and EDS mapping showing the conformal deposition of Ni and CN on the roughened surface of RCD-PyC.} (A to D) RCD-PyC electrode before surface modification. (E to H) Ni/RCD-PyC electrode. (I to L) CN/Ni/RCD-PyC electrode. M is the SEM image of CN/Ni/RCD-PyC and N, O and P are the corresponding EDS mapping images.}
    \label{fig.4}
\end{figure}
The morphology of the modified electrode surfaces was characterized using SEM spectroscopy as shown in Figure \ref{fig.4}.
SEM images of RCD-PyC reveal that pyrolysis of the 3D polymer has resulted in a uniform shrinkage of the structure by three times smaller than the initial volume.
Further, the higher-resolution SEM images show that RCD-PyC has a rough surface (Figure \ref{fig.4}B-D).
The volume reduction of the 3D polymer precursor is ascribed to degassing and following desorption of the produced volatile gases\cite{Sun2024InMetamaterials}.
Carbonization of the organic polymer is known to take place in a multi-step process\cite{Nakagawa1987StudiesChromatography}.
In general, the primary stage is dehydration, followed by the elimination of volatile gases. Carbonization and volume reduction occur above 500 °C, along with eliminating nitrogen, hydrogen, and oxygen\cite{Natu2018ShrinkageCarbonization}.
The pyrolysis atmosphere plays a crucial role in the shrinkage of the polymeric precursor.
In the case of vacuum pyrolysis, the low-pressure environment leads to the formation of volatile gas at a lower temperature and enhanced gas diffusion rate, compared to a nitrogen and argon atmosphere\cite{Jia2008EffectsAdsorption}.
\newline Using a simple electrodeposition process, the surface of the RCD-PyC electrode was uniformly deposited with Ni nanoparticles, as shown in Figure \ref{fig.4}E-H).
The Ni coating obtained by electrodeposition produced a globular-like morphology with considerable roughness.
This results in an overall higher surface area electrocatalyst.
The generally accepted mechanism for Ni electrodeposition in the Watts bath is as follows (Eq. (\ref{Eq. 1})-(\ref{Eq. 3}))\cite{Nasirpouri2013Refinement1,Saraby-Reintjes1984KineticsBaths,Abyaneh1981TheElectrocrystallisation,Razika2018MechanismNickel}:
 \begin{equation} \label{Eq. 1}
 \mathrm{{Ni^{2+}} + {X^{-}} \rightarrow {NiX^{+}}}
 \end{equation}
 \begin{equation} \label{Eq. 2}
 \mathrm{{NiX^{+}} + {e^{-}} \rightarrow {NiX_{ads}}}
 \end{equation}
 \begin{equation} \label{Eq. 3}
 \mathrm{{NiX_{ads}} + {e^{-}} \rightarrow {Ni} + {X^{-}}}
 \end{equation}
where \ch{X-} has been considered to be \ch{OH-} or \ch{Cl-}.
SEM images of Ni deposition cross-sections are given in \sfref{supp-fig:S2} to provide a detailed visualization of the distribution and morphology of the deposited layer. 

\begin{figure}[!ht] 
    \centering
    \includegraphics[width=\linewidth]{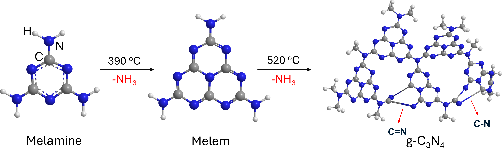}
    \caption{\textbf{Reaction pathway of CN formation from thermal condensation of melamine through CVD.}}
    \label{fig.5}
\end{figure}

Consequently, conformal deposition of carbon nitride was achieved via CVD. CVD is a highly conformal deposition technique, allowing uniform deposition of complex geometries and roughened surfaces\cite{Angus2008ChemicalDeposition}. CN was directly grown on the Ni/RCD-PyC electrode surface using chemical vapor deposition (CVD). Melamine served as the precursor for the conformal coating of CN on the nickel-electrodeposited 3D-printed carbon electrode within the CVD reaction furnace tube. Polymeric carbon nitrides form Melem at temperatures above 390°C, transforming into the CN network at temperatures exceeding 520°C\cite{Zhang2012PolymericConversion}. 
The proposed CVD mechanism for the formation of CN from melamine involves the thermal decomposition of melamine, which releases ammonia at each stage. This process results in the progressive formation of CN with an increasingly graphitic structure (Figure \ref{fig.5}).
Each step emphasises the loss of ammonia and the transformations of the precursor, illustrating the condensation process as follows (Eq. ~\ref{Eq. 4} and ~\ref{Eq. 5}:

Step 1 - Melamine to melem conversion:
\begin{equation} \label{Eq. 4}
    \mathrm{ 2 \,\ch{C3H6N6} \, ({melamine}) \rightarrow \ch{C6H6N10} \, ({melem})
     + 2\,\ch{NH3}}
 \end{equation}
 
Step 2 - Melem to polymeric carbon nitride conversion:
 \begin{equation} \label{Eq. 5}
     \mathrm{n\,\ch{C6H6N10} \, ({melem}) \rightarrow   2n \,\ch{C3N4} \, (\ch{CN})+ 2n\,\ch{NH3}}
 \end{equation}
 
The details of the CVD process are described in the Materials and Methods section. The CVD-grown CN thin films were characterized using SEM, EDS, XRD, XPS, Raman, and focused ion beam (FIB).
SEM (Figure \ref{fig.4}M) and EDS mapping images (Figure \ref{fig.4}N-P) depict that the further roughened surface of Ni/RCD-PyC electrode has been conformally coated with CN, whereas CN deposition resulted in uniform deposition, which is crucial for consistent electronic properties.
SEM images, including both top-view and cross-sectional views of CN deposited on silica wafers, are provided in \sfref{supp-fig:S3} to further illustrate the deposition characteristics and structural morphology. 
Additionally, lower-resolution SEM images of the electrodes are given in \sfref{supp-fig:S1} to offer a broader view of the material morphology. 
The annealing of the vacuum furnace in a controlled atmosphere enabled a uniform shrinkage of the 3D-printed polymeric structure.

The EDS spectra of the 3D electrodes at different stages of fabrication are given in \sfref{supp-fig:S4}.
EDS spectra of the CN/Ni/RCD-PyC reveal that CVD has successfully deposited CN without impurities.
Elemental mapping images of Ni/RCD-PyC and CN/Ni/RCD-PyC show that Ni and nitrogen are uniformly deposited on the RCD-PyC electrode surface.
Furthermore, elemental composition analysis of the electrocatalyst reveals the compositions of Ni, carbon and nitrogen.
To understand the composition of CN without the interference of carbon from RCD-PyC, the elemental composition of CN deposited on the silica wafer was analyzed (\sFref{supp-fig:S5}).
These data are consistent with the elemental composition corresponding to nitrogen-rich CN in the literature\cite{Mane2017HighlyGeneration}.

As demonstrated in \sfref[A]{supp-fig:S6}, XRD spectra of Ni coating on the ITO glass slide exhibited a typical diffraction pattern for single-phase face-centerd cubic (FCC) structures.
The crystallite size of the Ni nanoparticles was calculated using the Scherrer equation (Eq. (\ref{Eq. 6}))\cite{Shahroudi2024EnhancedElectrocatalyst}. 
\begin{equation} \label{Eq. 6}
    \mathrm{{D}={K\lambda}/W \, Cos\theta}
\end{equation}
where D represents the average crystallite size, K is the Scherrer constant, $\lambda$ is the wavelength of X-ray radiation, the $\theta$ denotes the X-ray diffraction angle, and W is obtained through converting the full width at half maximum (FWHM) from degree to radian according to Eq. (\ref{Eq. 7}).
\begin{equation} \label{Eq. 7}
    \mathrm{W={FWHM \times 3.1416/180}}
\end{equation}
The average crystallite size for the (111) plane of the pure Ni nanoparticles deposited on ITO was estimated to be about 27.5 nm. 
The X-ray photoelectron spectroscopy (XPS) technique was employed to analyze the chemical state and probable elemental compositions of the CN/Ni/RCD-PyC electrode.
The C1s spectrum (\sFref[A]{supp-fig:S7}\cite{Hellgren2016InterpretationXPS}) exhibited peaks around 285.45 and 288.27 eV corresponding to C-N and C=N, respectively, as indicated in Figure \ref{fig.5}.
The additional small peak at the 288 eV in the C1s spectrum was ascribed to the C-O bond\cite{Dementjev2000X-RayFilms}.
As shown in \sfref[B]{supp-fig:S7}, the peaks around 399.15 and 401.30 eV in the N1s spectrum can be designated as pyridinic N and graphitic N, respectively\cite{Wu2015TransformingElectrocatalysts}, confirming the presence of graphitic CN.
The additional small peak around 404.78 is due to the presence of terminal-\ch{NO2} groups\cite{Gu2014HierarchicallyDegradation}.
The presence of a peak around 852.17 eV in the Ni2p spectrum (\sFref[C]{supp-fig:S7}) confirms the presence of metallic nickel (Ni${^0}$) along with satellite peaks at 858.63 and 869.84 eV\cite{Dang2014EnhancingNm}.
However, in other regions, a small peak at 854.9 eV was detected, indicating the presence of trace amounts of Ni${^{2+}}$ species such as nickel oxide(NiO) and nickel hydroxide (\ch{Ni(OH)2}).
These Ni${^{2+}}$ species are likely formed due to surface exposure to the atmosphere, contributing to the observed heterogeneity.
The O1s spectrum, deconvoluted into two peaks around 532.28 and 534.23 eV represents C-O bonds and lattice water molecules, respectively.
However, the expected Ni-O peak was not resolved, likely due to its overlap with the stronger C-O signals and low intensity compared to the C-O contribution (\sFref[D]{supp-fig:S7})\cite{Ghosh2021NickelcobaltReaction}.

\begin{figure}
    \centering
    \includegraphics[width=\linewidth]{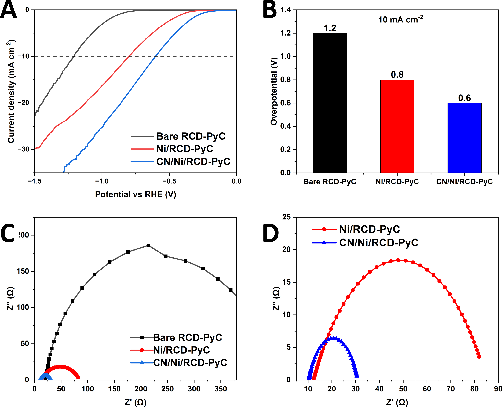}
    \caption {\textbf{Electrochemical performance analysis of electrodes in 0.5 M  \ch{H2SO4}:} (A) Linear sweep voltammetry depicting the HER overpotential of CN/Ni/RCD-PyC, Ni/RCD-PyC and RCD-PyC electrodes in 0.5 M  \ch{H2SO4}.
    (B) Bar graph depicting overpotentials at 10~\mAcmsquared{} obtained from corresponding LSV curves.
    (C) EIS spectra of all electrodes and modified electrodes, respectively, measured at overpotential values at a frequency range of 0.1 Hz to 10 kHz.
    (D) Zoomed-in view of EIS spectra, showing closer comparison between Ni/RCD-PyC and CN/Ni/RCD-PyC electrodes.}
    \label{fig.6}
\end{figure}

Raman spectroscopy was performed to analyse the chemical structure of the electrocatalysts.
Raman spectra of the RCD-PyC modified with CN, before and after nickel electrodeposition, were compared.
The RCD-PyC modified with nickel and CN has two peaks at 1367 cm${^{-1}}$ and 1590 cm${^{-1}}$, corresponding to the D and G peaks of CN, respectively.
This spectrum is also composed of an additional peak at $\sim$2197 cm$^{-1}$ which corresponds to Ni nanoparticles on the surface of the RCD-PyC electrode (Figure \ref{fig.6}C)\cite{Yousef2017FacileCells}.
Furthermore, an enhancement in the intensity of the peaks indicates that the presence of nickel has improved CN deposition. 
Supplementary Figure \ref{fig.6}B represents the XRD pattern of CN deposited on a quartz substrate.
The strong peak at 27.38\textdegree{} can be depicted as the diffraction peak corresponding to the (002) lattice plane of CN.
This is a characteristic of periodic interlayer stacking along the C-axis of CN.
\section{Electrochemical Characterisation}
\begin{figure}
    \centering
    \includegraphics[width=\linewidth]{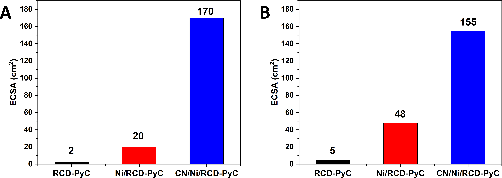}
    \caption{\textbf{Comparison of \ac{ECSA} of electrodes after sequential surface modification:} Bar graphs for the \ac{ECSA} for different electrodes in (A) acidic and (B) alkaline electrolytes. }
    \label{fig:7_ECSA_comparison}
\end{figure}

First, the HER electrocatalytic activity of the electrodes was tested using a three-electrode electrochemical cell in 0.5 M aq. {H$_{2}$SO${_4}$} solution as the electrolyte. 
The electrochemical activity of the prepared electrode was studied using linear sweep voltammetry (LSV), at 25\textdegree{}C in the potential range from 0 to -1.3 V (vs Hg/\ch{Hg2Cl2}), at a scan rate of 100 mV s$^{-1}$ (Figure \ref{fig.6}A).
The electrochemical activity of the bare 3D carbon electrode and the nickel electroplated 3D electrode was also analysed.
The bare RCD-PyC electrode showed insignificant electrocatalytic activity, requiring a high overpotential of 1.2~V to achieve a current density of 10~\mAcmsquared{}.
CN/Ni/RCD-PyC required an overpotential of -0.6~V while Ni/RCD-PyC required an onset potential of -0.8~V at 10~\mAcmsquared{} current density.
The \acp{ECSA} of each electrode were analysed using \acf{CV} at different scan rates to obtain the double layer capacitance (C\textsubscript{dl}) as shown in \sfref{supp-fig:S9_ECSA_H2SO4}.
The calculated \ac{ECSA} values are shown in the bar graph (Figure \ref{fig:7_ECSA_comparison}A).
In acidic media, the ECSA was measured as 170~cm$^2$ for CN/Ni/RCD-PyC electrode, 20~cm$^2$ for Ni/RCD-PyC and 2~cm$^2$ for bare RCD-PyC, highlighting the substantial improvement in surface area attributable to CN/Ni/RCD-PyC due to improved stability and electrical conductivity (Figure \ref{fig.6}).
In acidic conditions, the enhancement factor (ECSA:GSA) ratios are RCD-PyC : Ni/RCD-PyC = \ECSAoverGSAacidicBare{}:\ECSAoverGSAacidicNi{} and RCD-PyC : CN/Ni/RCD-PyC = \ECSAoverGSAacidicBare{}:\ECSAoverGSAacidicCN{}.
This suggests that the ECSA is over 80 times greater than the GSA for CN/Ni/RCD-PyC.
This is attributed to the increased electrochemical activity due to the spillover effect of CN on hydrogen adsorption.
Electrochemical impedance spectroscopy (EIS) of the samples was performed at the overpotential values and a sinusoidal perturbation amplitude of 1 mV.
Figure \ref{fig.6} (C and D) illustrates the corresponding Nyquist plots of the RCD-PyC electrode modified with Ni and CN in 0.5 M aq. {H$_{2}$SO${_4}$}.
All the curves are fitted by a simplified equivalent circuit as shown in \sfref{supp-fig:S11_EIS_equivalent_circuit}.
The equivalent circuit consists of a solution resistance (R$_{S}$), corresponding to the contact resistance between the electrolyte solution and the electrode, a constant phase element (CPE) and charge transfer resistance (R\textsubscript{CT}).
All samples exhibited Nyquist plots in the form of a semicircle, indicating the resistance to charge transfer\cite{Elumalai2002KineticsStudies}.
The diameter of the semicircle decreases after consecutive surface modifications, indicating a reduction in the charge-transfer resistance.
The smallest EIS curve diameter validates the increased charge transfer and corrosion resistance properties of RCD-PyC after surface modification with CN\cite{Fayyad2018SynthesisNanocomposites}.
This is attributed to better charge transfer and chemical interaction between the RCD-PyC electrode surface and electrocatalytic materials.
The values for each electrode in both electrolytes corresponding to R\textsubscript{S}, CPE and R\textsubscript{CT} are provided in \stref{supp-tab:S3_EIS_fit_parameters}.
\begin{figure}
    \centering
    \includegraphics[width=\linewidth]{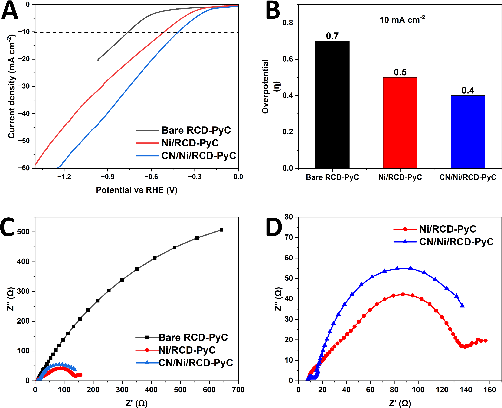}
    \caption{\textbf{Electrochemical performance analysis of electrodes in 1 M KOH.} (A) Linear sweep voltammetry depicting the HER overpotential of CN/Ni/RCD-PyC, Ni/RCD-PyC and RCD-PyC electrodes in 0.5 M  \ch{KOH}. (B) Bar graph depicting overpotentials at 10~\mAcmsquared{}, obtained from corresponding LSV curves. (C) EIS spectra of all electrodes and modified electrodes, respectively, were measured at overpotential values at a frequency range of 0.1 Hz to 100k Hz. (D) Zoomed-in view of EIS spectra, showing closer comparison between Ni/RCD-PyC and CN/Ni/RCD-PyC electrodes.}
    \label{fig:8_LSV-EIS-KOH}
\end{figure}

The HER electrocatalytic performance of the electrodes was further analysed and compared in 1M aq. KOH solution.
Like the acidic electrolyte, the untreated RCD-PyC electrodes exhibited huge overpotentials, representing weak electrocatalytic activity.
After deposition of the electrocatalysts, the Ni/RCD-PyC and CN/Ni/RCD-PyC required overpotentials of -0.5 V and -0.4 V to achieve a current density of 10~\mAcmsquared{} (Figure \ref{fig:8_LSV-EIS-KOH}, A and B).
This is attributed to the efficient utilization of electrode surface area for the deposition and subsequent electrochemical activity.
Moreover, the effect of CN on the electrocatalytic effect was analysed using EIS (Figure \ref{fig:8_LSV-EIS-KOH}, C and D).
Results suggest that the R\textsubscript{CT} has largely reduced upon subsequent deposition of nickel, followed by CN.
The calculated ECSA values are presented in a bar graph in Figure \ref{fig:7_ECSA_comparison}B.
As depicted, the ECSA demonstrates a marked improvement when the RCD-PyC is coated sequentially with Ni and then CN, increasing from 5.35 cm${^{2}}$ to 155 cm${^{2}}$.
Under alkaline conditions, the ratios are RCD-PyC : Ni/RCD-PyC = \ECSAoverGSAalkalineBare{}:\ECSAoverGSAalkalineNi{} and RCD-PyC: CN/Ni/RCD-PyC = \ECSAoverGSAalkalineBare{}:\ECSAoverGSAalkalineCN{}, indicating that the ECSA is over 30 times larger than the GSA for CN/Ni/RCD-PyC compared to bare RCD-PyC.
This enhancement is attributed to the synergistic interaction of active materials leading to enhanced catalytic activity, corrosion resistance, and improved active sites.

\begin{figure}[h]
    \centering
    \includegraphics[width=\linewidth]{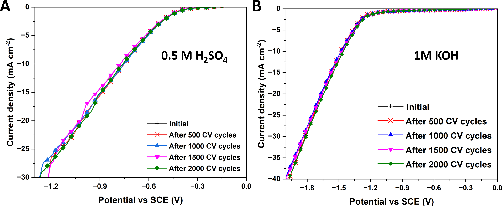}
    \caption{
    \textbf{Stability analysis of CN/Ni/RCD-PyC in acidic and alkaline electrolytes.} Linear Sweep Voltammetry (LSV) after each 500 CV cycles until 2000 CV cyclic stability test in (A) acidic and (B) alkaline electrolytes, demonstrating the catalytic performance over time and showing cyclic stability.
    }
    \label{fig.9}
\end{figure}

The stability of the electrocatalytic system was evaluated using CV in both acidic and alkaline media through 2000 CV cycles and shows exceptional stability, as shown in Figure \ref{fig.9}.
This is attributed to catalytic activation resulting from increased surface roughening.
This was confirmed using SEM imaging conducted after cyclic stability analysis (\sFref{supp-fig:S12_SEM_cyclic_stability}).
The LSV curves indicating the cyclic stability over time have been shown in Figure \ref{fig.9}. 

\section{Conclusion}
By integrating 3D-printing and a unique modification strategy, we have presented an innovative framework for catalyst structure high volume-to-surface area enhancement ratio and improved material stability.
While pyrolysis of 3D-printing polymer offers inherently porous surface morphology, the ECSA of the electrocatalyst was tuned by optimizing the conditions and precursor composition for electrodeposition and CVD. 
This study explored the fabrication and performance of 3D-PyC with a unique RCD structure, derived from SLA and vacuum pyrolysis for electrocatalytic HER application in acidic and alkaline media.
The electrode was further modified with active electrocatalyst materials via electrodeposition and CVD.
We demonstrated this method by developing a surface-functionalized 3D-printed carbon electrode with an enhancement ratio of 1:80 from the GSA to the ECSA. 
The superior HER catalytic behaviour of CN/Ni/RCD-PyC in both acidic and alkaline media was demonstrated. Importantly, the durability of the catalytic system was analysed using 2000 CV cycles.
Modifying the Ni/RCD-PyC electrode with CN enhanced the ECSA and, hence, HER activity in both acidic and alkaline media.
The enhancement factors (ECSA:GSA), indicating the extent to which the \ac{ECSA} is increased relative to the \ac{GSA} for different materials or conditions, are listed in \stref{supp-tab:S2_ECSA_over_GSA}.
This is attributed to improved chemical stability and an increased water dissociation rate in acidic and alkaline electrolytes, respectively.
This work presents several avenues for future research. The approach outlined in this study should be considered for different geometries and other catalysts such as titanium and molybdenum.
Further investigation should consider electrodeposition and CVD based on various compositions, depositing parameters, and similar deposition techniques.
This research provides insights into developing and analyzing RCD-PyC electrodes for alkaline and acidic HER.
This study provides a fresh perspective on 3D electrode development, offering valuable insights into material design for efficient and robust HER electrocatalysts, and contributing to the advancement of sustainable hydrogen production technologies.

\ack
This work was supported by the Energy Futures Seed Fund and Research Development Fund (RDF) Studentship Scheme at Northumbria University and the Engineering and Physical Sciences Research Council (EPSRC) grants EP/N00762X/1, EP/W022931/1, EP/V040030/1,
EP/Y003551/1, and EP/Y016440/1.
We also thank Pietro Maiello and Rebecca Payne from the Northumbria Materials Characterization Laboratory (NMCL) for their invaluable assistance with SEM, and EDS analysis. We also thank Jake Sheriff from the National ESCA and XPS Users' Service (NEXUS) at Newcastle University for his help with XPS measurements.

\section*{Data availability statement}

All data needed to evaluate the conclusions in the paper are present
in the paper and/or the supplementary materials.
Additional data related to this paper may be requested from the authors.

In addition, the data and script used to create \sfref{supp-fig:S8_electrodeposition_SH_model_fit} are made available via the following Zenodo repository\cite{meethale_palakkool_2025_15295929}:
\emph{"Scharifker-Hills model fits of experimental measurements of current-time transients during Nickel electrodeposition on pyrolitic carbon at four different applied potential values: -1.1, -1.2, -1.3 and -1.4 V vs Ag/AgCl."}, \url{https://doi.org/10.5281/zenodo.15295929}.
The datasets are available under the terms of the Creative Commons  Attribution 4.0 International license.
The source code used to generate the figure is available under the GPL 3.0 license.

\appendix
\section*{Appendix}
\section{Materials}
The commercial photopolymer High Temp V2 was acquired from Formlabs.
Nickel(II) chloride hexahydrate (\ch{NiCl2.H2{O}}), Nickel(II) sulfate hexahydrate (\ch{NiSO4.H2{O}}), boric acid (\ch{H3BO3}), potassium hydroxide (KOH), sulphuric acid (\ch{H2SO4}) were obtained from Sigma-Aldrich.
Melamine (\ch{C3N6H6}, 99\% pure) was obtained from Thermo Scientific Chemicals. 

\section{Synthesis of RCD-PyC}\label{sec:method_fabrication}
The Formlabs High-Temperature V2 resin was chosen as the precursor photopolymerizable resin, as it has been previously reported to maintain structural integration under heat treatment\cite{Rezaei2020HighlyTechnology}. The 3D structure was designed using Blender and exported as an SLA file (.stl) into slicing software for the Formlabs 3D printer, the Preform. Models were 3D-printed via a Formlabs 2 3D printer.
The structures were directly printed on the build platform at a resolution of 25 µm.
Dimensions of the electrode geometry were optimised as a cylindrical geometry, with a diameter of 17.4 mm, and a height of 6 mm.
The 3D-printed structures were washed with isopropyl alcohol using Form Wash to remove any remaining cross-linked residual resins.
Form Wash works based on a magnetically coupled impeller-based agitation method.
The 3D-printed structures were removed from the 3D printer's build platform and subjected to Form Wash for 6 min.
To ensure the complete curing of liquid resins on the 3D-printed structures, Form Cure was conducted for 60 minutes at 60 °C.

The 3D-printed polymers were carbonized through a three-step pyrolysis process; specimens were first pyrolyzed at 400 °C for 4 hours, followed by pyrolysis at 1000 °C for 4 hours and finally cooled down to room temperature for 4 hours.
The process was carried out at a constant ramping rate of 3 $^{\circ}$C/min under vacuum conditions (Figure \ref{fig:3}A).

\section{Synthesis of Ni/RCD-PyC}
The electroplating solution was prepared using deionized water obtained from Millipore.
The solution bath consisted of 24 g of nickel sulphate, 4 g of nickel chloride, and 4.5 g of boric acid.
All solutions were prepared in 100 mL volume.
During the electrodeposition process, the solution was stirred with a magnetic stirrer at a constant speed of 600 rpm at 70 $^{\circ}$C (Figure \ref{fig:3}A).
The electrodeposition process was performed for 1800 s for each sample.
The electrodeposition was carried out using a three-electrode electrochemical cell, consisting of a 3D electrode, a Pt wire, and an Ag/AgCl electrode as the working electrode (WE), counter electrode (CE), and reference electrode (RE), respectively.
Electroplating was carried out using chronoamperometry techniques at a potential of –1.3 V.
The current-transient analysis of the theoretical formulation was performed using Python 3.12 (see \sfref{supp-fig:S8_electrodeposition_SH_model_fit} and \cite{meethale_palakkool_2025_15295929}).

\section{Synthesis of CN/Ni/RCD-PyC}

CN was directly grown on the Ni/RCD-PyC electrode surface to produce a conformal coating through CVD. In our CVD conformal coating design, nickel-electrodeposited 3D-printed carbon electrodes were placed on the substrate holder and maintained at temperatures between 500°C and 650°C (typically 550°C) for 2 hours, with a ramping rate of 20°C/min. Meanwhile, 800 mg of melamine was placed in a silica crucible and heated to 350°C using an external heating band. The melamine vapor was then delivered downstream to the substrate via argon gas at a 50 mL/min flow rate, controlled by a mass flow controller (MFC). The deposition process was conducted at 950 mbar, controlled by a Vacuubrand MV 10C NT Vario pump with a pressure controller.
Figure \ref{fig:3}A illustrates the experimental setup and a schematic depiction of the conformally CVD-grown CN thin films.

\section{Characterization}
Considering the RCD-PyC electrode's rough surface made it unsuitable for x-ray diffraction spectroscopy (XRD) analysis, the modification processes were repeated on a silica wafer and quartz substrate for CVD and an ITO-coated glass slide for nickel electrodeposition.
Material morphology was obtained using scanning electron microscopy (SEM) using a field emission TESCAN MIRA 3 electron microscope with a large chamber.
The elemental composition was analyzed via energy-dispersive X-ray diffraction spectroscopy (EDS) using an attached in-beam detector. FIB cutting and cross-section analysis were conducted using FEI Helios NanoLab 600 with a Kleindeick micromanipulator for platinum deposition and force measurement. Crystallographic analysis was performed via XRD with Rigaku SmartLab SE using a Cu K$\alpha$ radiation source, $\lambda$= 0.15406 nm. Height alignment was performed before each acquisition. XPS analysis was performed using a K-Alpha instrument (Thermo Scientific) equipped with a monochromatic X-ray source (Al, K$\alpha$ = 1486.6 eV). Spectra were acquired in Fixed Analyser Transmission (FAT) mode with a 4.2 eV work function.
Raman spectroscopy was performed to further probe the structure of deposited materials, with a Renishaw InVia Raman spectrometer equipped with a 532 nm laser source, with the excitation source calibrated with Si control reference. 

\section{Electrochemical Measurements}
All \ac{HER} electrocatalytic testing was conducted on a typical three-electrode electrochemical cell using the Metrohm Autolab potentiostat and NOVA 2.1.3 software.
A standard three-electrode electrolyser with 1M \ch{KOH} and 0.5 M \ch{H2SO4} was used for measurements.
A platinum wire and the saturated calomel electrode (Hg/\ch{Hg2Cl2}, 3M KCl) were used as counter electrode (CE) and reference electrode (RE), respectively.
The \ac{HER} performance of the electrodes was measured by linear sweep voltammetry (LSV) and electrochemical impedance spectroscopy (EIS).
Additionally, the ECSA was measured using CV in a non-faradaic (capacitive) region.
LSV curves were measured at a scan rate of 100~mV\textperiodcentered{}s$^{-1}$.
The potentials obtained were converted to reversible hydrogen electrodes using the Nernst equation (Eq. (\ref{Eq. 8})):  
\begin{equation} \label{Eq. 8}
\mathrm{E{_{RHE}} = E{_{o}} + E{_{Hg/{Hg}_{2}Cl_{2}}} + 0.059 * pH }
\end{equation} 
where E${_{RHE}}$ is the potential against reversible hydrogen electrode, E${_{o}}$ is the potential obtained, and E${_{Hg/Hg2Cl2}}$ is the standard potential of Hg/\ch{Hg2Cl2} electrode (+0.241 V vs RHE).
The electrode was washed multiple times with ethanol and deionized (DI) water before electrochemical analysis. Before electrochemical measurements, the working electrode (WE) was immersed in an electrolyte solution for at least 30 minutes to allow complete wetting of the 3D electrode surface\cite{Anantharaj2022DosElectrocatalysts}.
They were also electrochemically activated with 20 cycles of cyclic voltammetry (CV), at a scan rate of 20 mV s$^{-1}$.
The \acp{GSA} of the 3D electrodes were measured using SEM imaging and the creation of an analogue CAD file, as explained in Section~\ref{sec:geometry}.
This area was used to determine the \ac{ECSA}.
This was done by conducting CV at scan rates of 10~mV\textperiodcentered{}s$^{-1}$ to 80~mV\textperiodcentered{}s$^{-1}$ at the capacitive region to determine the electrochemical double layer capacitance C\textsubscript{dl}.
The relationship between current (I) and scan rate ($\nu$) is given by Eq. (\ref{Eq. 9}):
\begin{equation} \label{Eq. 9}
\mathrm{I = {C_{dl}} \times {\nu}}
\end{equation}
where I is the current at a specific potential and $\nu$ is the scan rate (mV s$^{-1}$).
The C\textsubscript{dl} of the electrode is represented by the slope of this graph.
The ECSA was calculated using the obtained C\textsubscript{dl} values based on Eq. (\ref{Eq. 10}):
\begin{equation} \label{Eq. 10}
\mathrm{ECSA = \frac{C_{dl}}{C_{s}}}
\end{equation}
The specific capacitance (C\textsubscript{s}) values were taken as 35 and 40 \textmu F cm$^{-2}$ for acidic and alkaline electrolytes, respectively\cite{Connor2020TheCatalyst}.
EIS was performed for a frequency range of 100 kHz to 0.1 Hz at an AC amplitude of 1 mV at overpotential values.
The cyclic stability analysis was conducted at potential ranges -0.6 V to -0.9 V and -1.6 to -1.9 V in acidic and alkaline media, respectively, over 2000 CV cycles at a scan rate of 100~mV\textperiodcentered{}s$^{-1}$.

\printbibliography 

\end{document}


\title[Supplementary Materials]{Supplementary Materials for\\ \longTitle{}}

\author{
Nadira Meethale Palakkool$^{1}$,
Mike P. C. Taverne$^{1,2,\ast}$,
Owen G. Bell$^{1}$,
Christopher P. Jones$^{3}$,
Jonathan D. Mar$^{4}$,
Duc Tam Ho$^{1}$,
Vincent Barrioz$^{1}$,
Yongtao Qu$^{1}$,
Zhong Ren$^{5}$,
Chung-Che Huang$^{6,\ast}$,
Ying-Lung Daniel Ho$^{1,2,\ast}$
}

\address{$^{1}$ School of Engineering, Physics and Mathematics, Northumbria University, UK}
\address{$^{2}$ School of Electrical, Electronic and Mechanical Engineering, University of Bristol, Bristol, UK}
\address{$^{3}$ Interface Analysis Centre, H.H. Wills Physics Laboratory, University of Bristol, Bristol, UK}
\address{$^{4}$ School of Mathematics, Statistics and Physics, Newcastle University, Newcastle upon Tyne, UK}
\address{$^{5}$ Oxford Instruments Plasma Technology, Govier Way, Severn Beach, Bristol, UK}
\address{$^{6}$ School of Electronics and Computer Science, University of Southampton, Southampton, UK}

\ead{
mike.taverne@northumbria.ac.uk;
cch@soton.ac.uk;
daniel.ho@northumbria.ac.uk
}

\vspace{10pt}
\begin{indented}
\item[]\CustomDate{}
\end{indented}


\clearpage{}
\section*{SEM Images}

\begin{figure}[ht!]
    \centering
    \includegraphics[width=\linewidth]{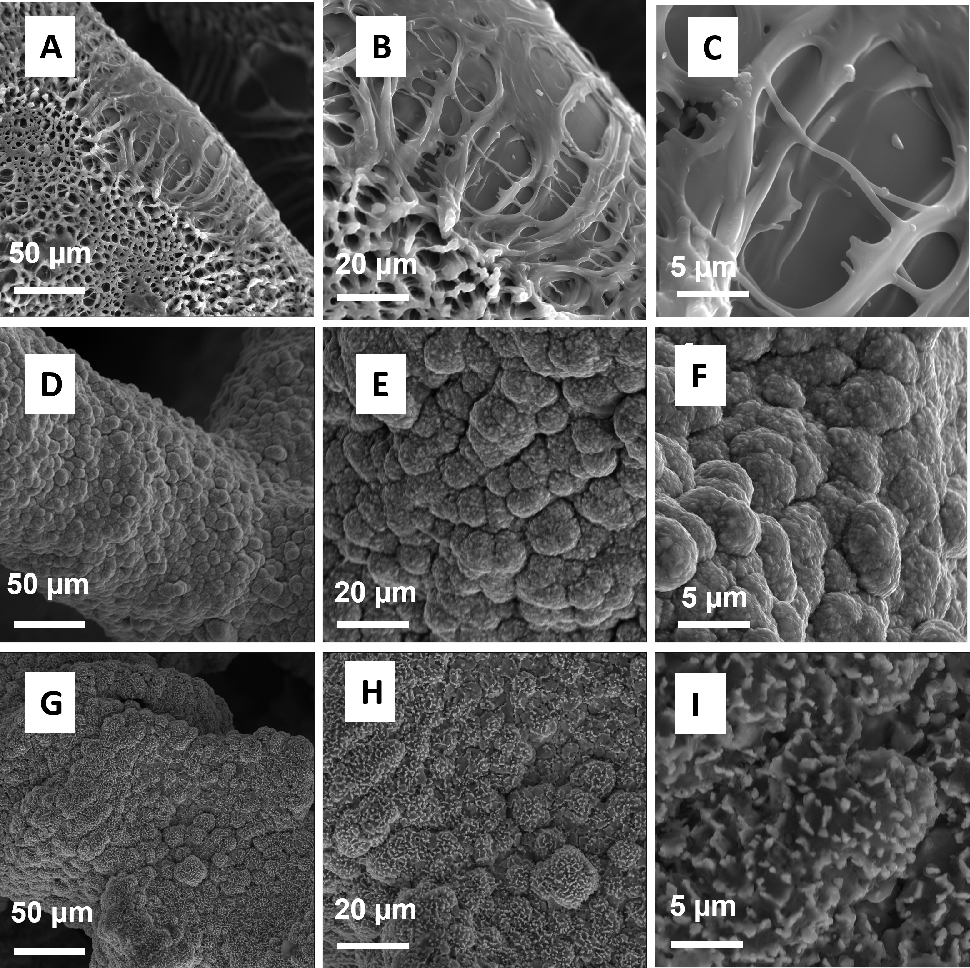}
    \caption{\textbf{High-resolution SEM Images: (A to C) Unprocessed RCD-PyC electrode. (D to F) RCD-PyC electrode modified with nickel and (G to I) RCD-PyC electrode modified with Ni, followed by CN}}
    \label{fig:S1}
\end{figure}

\newpage
\begin{figure}[ht!]
    \centering
    \includegraphics[width=\linewidth]{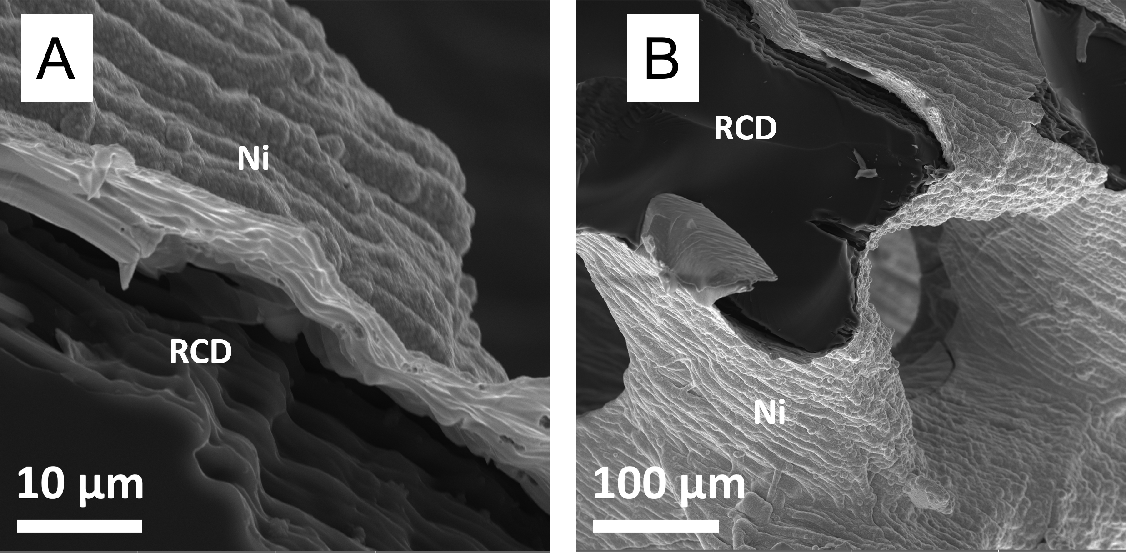}
    \caption{\textbf{SEM Images of cross-sectional view on nickel deposited on RCD-PyC electrode.}}
    \label{fig:S2}
\end{figure}

\begin{figure}[!ht]
    \centering
    \includegraphics[width=\linewidth]{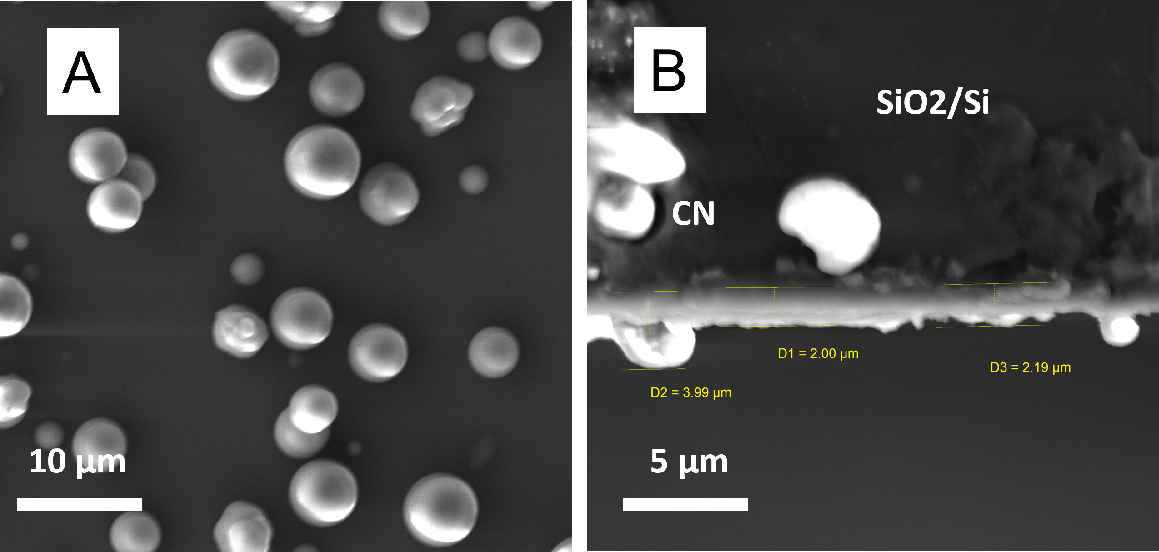}
    \caption{\textbf{SEM images of CN deposited on a silicon wafer with 300 nm \ch{SiO2}:} (A) top-view and (B) cross-section.}
    \label{fig:S3}
\end{figure}

\newpage
\section*{EDS Data}


\begin{figure}[!ht]
    \centering
    \includegraphics[width=\linewidth]{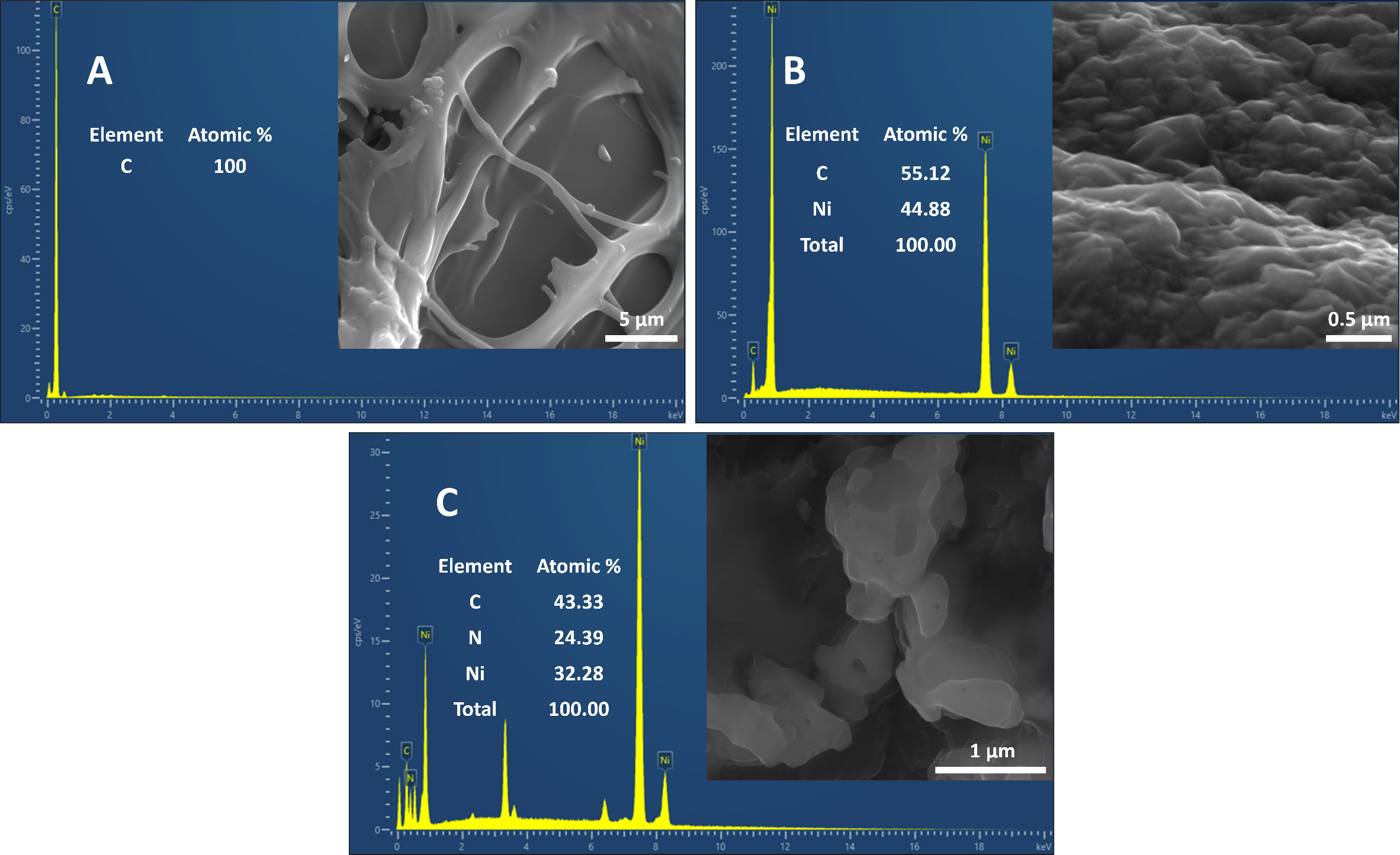}
    \caption{\textbf{EDS analysis of the electrodes:} The EDS spectra of (A) RCD-PyC, (B)Ni/RCD-PyC and (C) CN/Ni/RCD-PyC electrodes confirm the elemental composition at each stage of surface modification.}
    \label{fig:S4}
\end{figure}

\newcommand{\figSfourHeight}{0.30\linewidth}
\begin{figure}[!ht]
    \centering
    \includegraphics[width=0.7\linewidth]{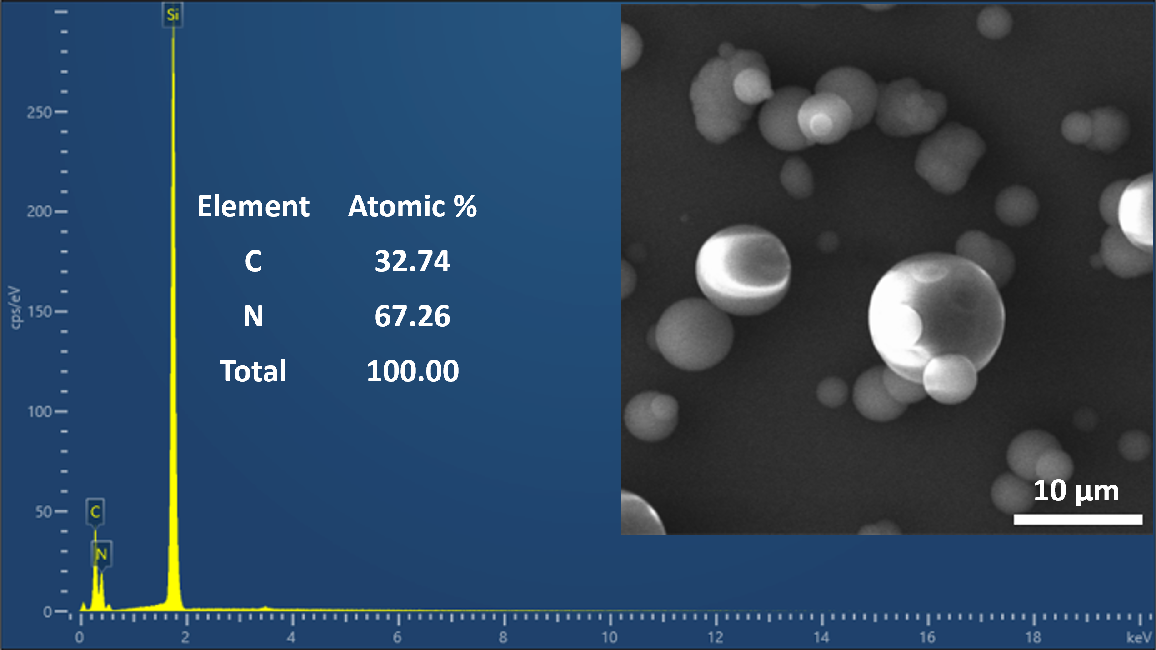} 
    \caption{\textbf{The EDS spectra of CN deposited on a silicon wafer with 300 nm \ch{SiO2} substrate.}}
    \label{fig:S5}
\end{figure}

\begin{figure}
    \centering
    \includegraphics[width=\linewidth]{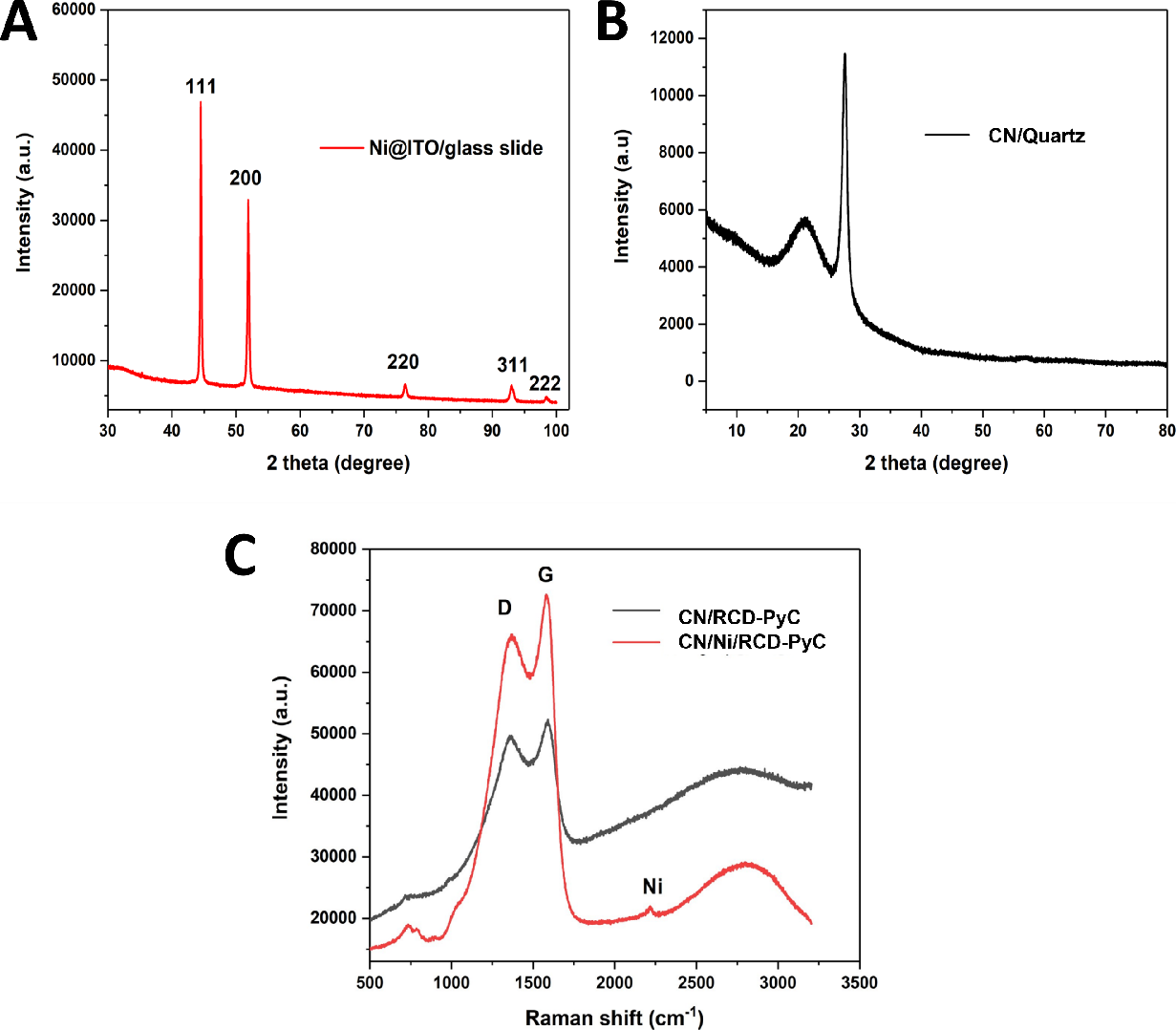}
    \caption{\textbf{Material characterisation on alternative substrates.}
    (A and B) XRD spectra of Ni electrodeposited on ITO/glass slide and CN deposited on quartz substrate, respectively. (C) Raman spectra of CN deposited on RCD-PyC electrode before and after Ni electrodeposition}
    \label{fig:S6}
\end{figure}

\begin{figure}[!ht]
    \centering
    \includegraphics[width=\linewidth]{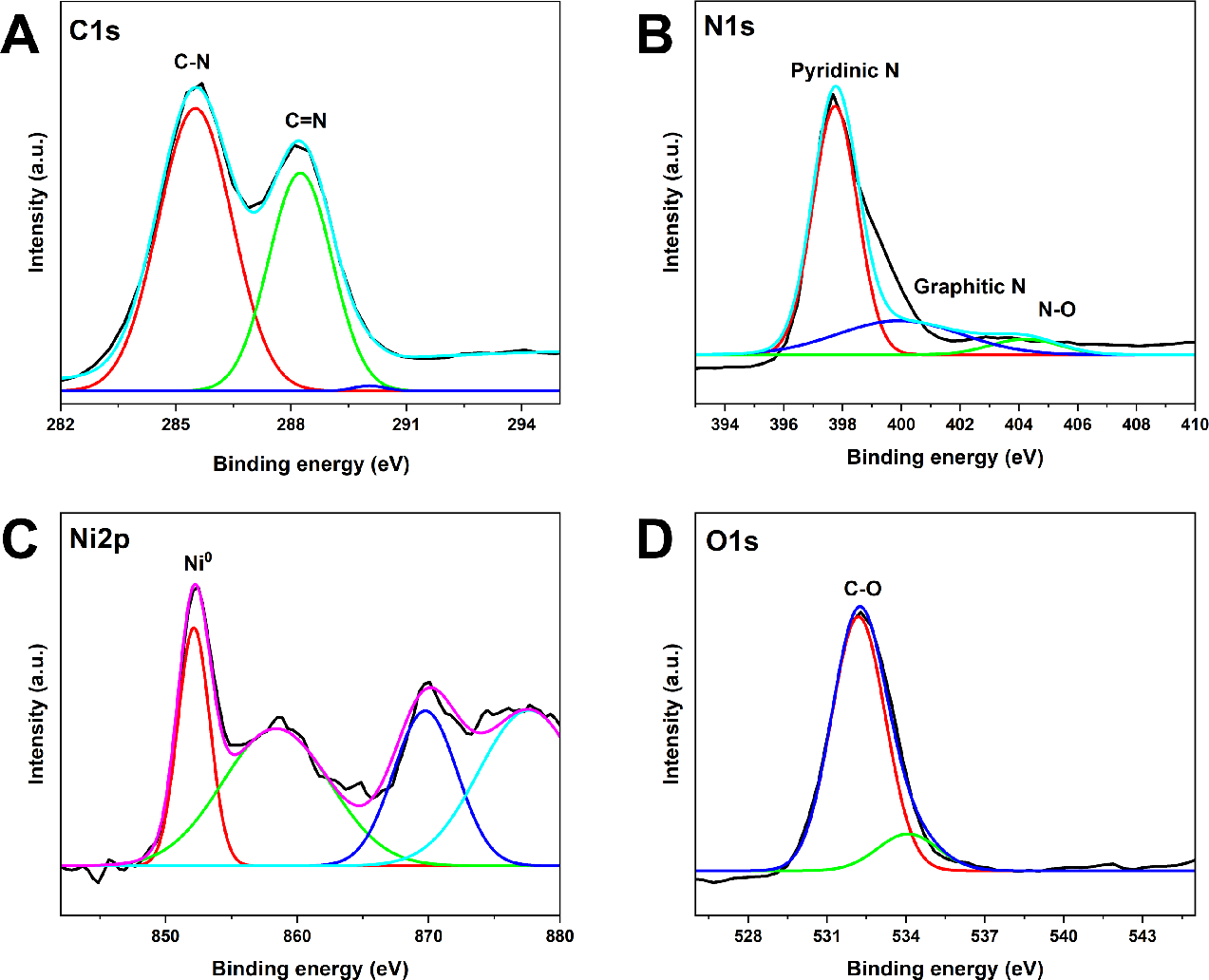}
    \caption{\textbf{The XPS spectra of CN/Ni/RCD-PyC electrode with narrow scan of (A) C1s, (B) N1s, (C) Ni2p and (D) O1s.}}
    \label{fig:S7}
\end{figure}

\clearpage{}

\section*{Theoretical Analysis of Electrodeposition}
\begin{figure}
    \centering
    \includegraphics[width=\linewidth]{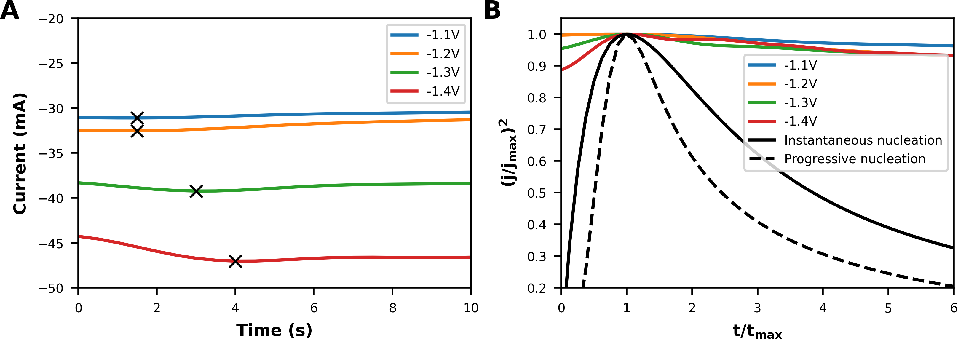}
    \caption{\textbf{Effect of deposition potential on the electrodeposition.} Effect of deposition potential on chronoamperometry of Ni electrodeposition process using Watts' bath.
    (A) Potentiostatic current-time transient curves.
    (B) Dimensionless plots of normalized current-time transient curves, as well as the fitted curves created using the Scharifker-Hill model using instantaneous nucleation (solid black line, Eq. S(\ref{eq:SH_instantaneous})) and progressive nucleation (dashed black line, Eq. S(\ref{eq:SH_progressive})).
    The black crosses in (A) indicate the position of the minima ($t_{max}$, $j_{max}$) used to normalize the curves in (B).
    The data and script used to create these figures are available via Zenodo\cite{meethale_palakkool_2025_15295929}.
    }
\label{fig:S8_electrodeposition_SH_model_fit}
\end{figure}

Chronoamperometry can be used as a deposition technique and a diagnostic tool to observe the formation and growth of stable nucleons\cite{Allam2018ElectrodepositionElectrolysis}.
The mechanism of nucleation was analysed using the Scharifker Hills model.
This model proposes two possible routes for the electrochemical nucleation process; instantaneous and progressive nucleation\cite{Scharifker1983TheoreticalNucleation}. Instantaneous nucleation can be described by Eq. S(\ref{eq:SH_instantaneous}):
\begin{equation} \label{eq:SH_instantaneous}
    \left(\frac{j}{j_{max}}\right)^2
    =
    \frac{1.9542}{t/t_{max}}
    \left\{
        1
        -
        exp\left[
            -1.2564 \left(\frac{t}{t_{max}}\right)
        \right]
    \right\}^2
\end{equation}

whereas progressive nucleation can be described as Eq. S(\ref{eq:SH_progressive}):

\begin{equation} \label{eq:SH_progressive}
    \left(\frac{j}{j_{max}}\right)^2
    =
    \frac{1.2254}{t/t_{max}}
    \left\{
        1
        -
        exp\left[
            -2.3367 \left(\frac{t}{t_{max}}\right)^2
        \right]
    \right\}^2
\end{equation}
where $j_{max}$ and $t_{max}$ represent the maximum current and time at which the maximum current occurs, respectively. Figure \ref{fig:S8_electrodeposition_SH_model_fit}A represents the experimental measurements of current-time transients at four different applied potential values, -1.1, -1.2, -1.3 and -1.4 V vs Ag/AgCl.
The sudden increase in current at the beginning is attributed to the double-layer charge formation. The current increases until it reaches the maximum current $j_{max}$, corresponding to the maximum time $t_{max}$ required for the nuclei growth on the electrode surface\cite{Torres2020ElectrochemicalIons}. 
Figure \ref{fig:S8_electrodeposition_SH_model_fit}A shows that electrodeposition follows instantaneous nucleation from a deposition potential of -1.1 V to -1.4 V.
The current value then increases and becomes constant, representing the diffusion-controlled nucleation process.
Figure \ref{fig:S8_electrodeposition_SH_model_fit}B shows that the current-time transients at different potentials follow similar shapes, suggesting a consistent mechanism.
However, increasing applied potential increases $j_{max}$ and correspondingly reduces $t_{max}$.
This shows that a larger applied potential will improve the electrocrystallization process by enhancing the surface concentration of metal ions.
Furthermore, dimensionless current-time transients were obtained using experimental data according to Eq. S(\ref{eq:SH_instantaneous}) and S(\ref{eq:SH_progressive}).

\clearpage
\section*{ECSA Calculation}
\begin{figure}[!ht]
    \centering
\includegraphics[width=\linewidth]{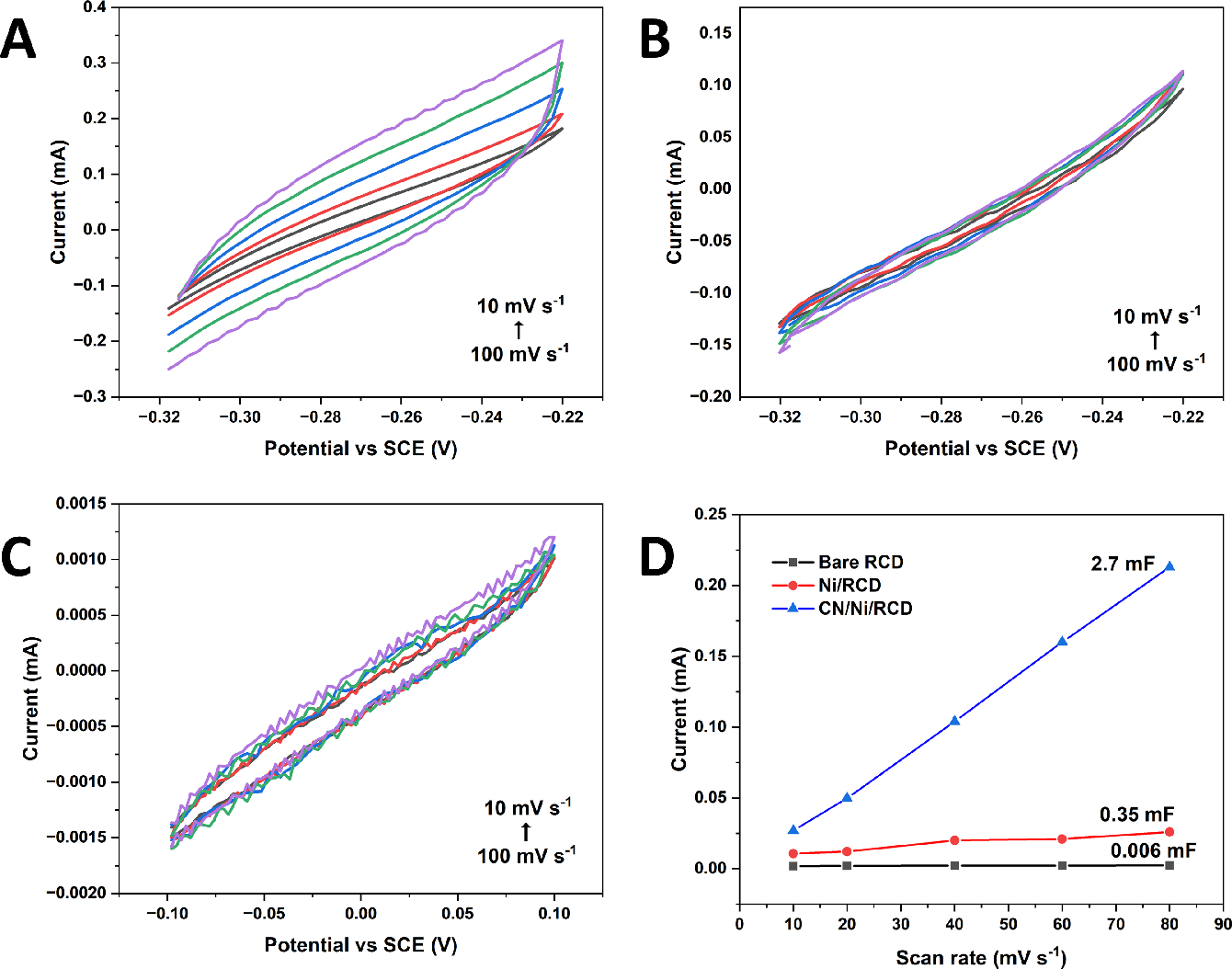}
    \caption{\textbf{ECSA measurements of electrocatalysts in 0.5 M \ch{H2SO4}.}
    Electrochemical cyclic voltammetry scans recorded for (A) CN/Ni/RCD-PyC, (B) Ni/RCD-PyC and (C) Bare RCD-PyC. (D) Linear fitting of capacitive current versus scan rates of CV measurements.}
    \label{fig:S9_ECSA_H2SO4}
\end{figure}

\clearpage
\begin{figure}[h!]
    \centering
\includegraphics[width=\linewidth]{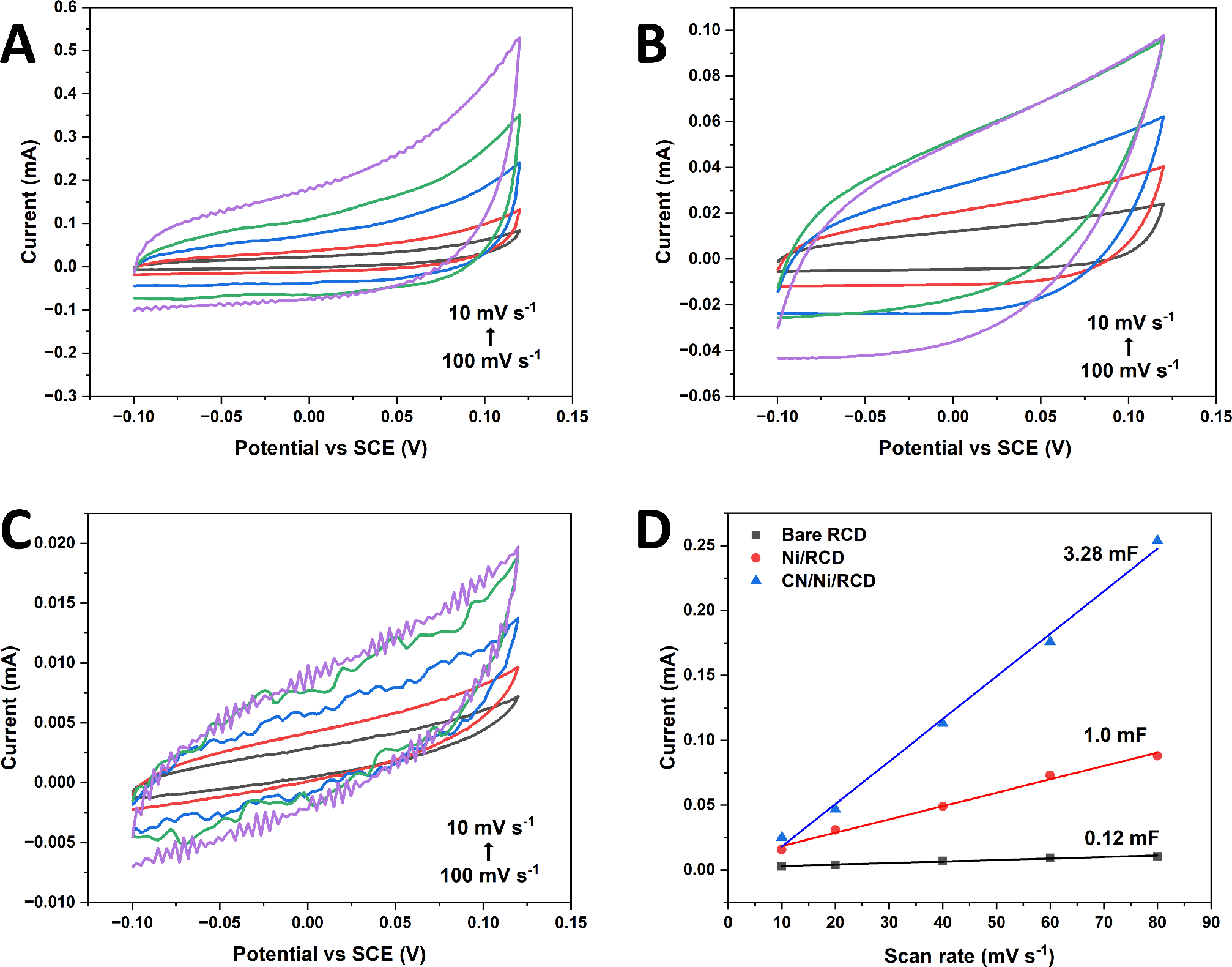}
    \caption{\textbf{ECSA measurements of electrocatalysts in 1 M KOH.} Electrochemical cyclic voltammetry scans
recorded for (A) CN/Ni/RCD-PyC, (B) Ni/RCD-PyC and (C) Bare RCD-PyC. (D) Linear fitting of capacitive current versus scan rates of CV measurements. }
    \label{fig:S10_ECSA_KOH}
\end{figure}

\clearpage
\section*{Surface Area Enhancement Factors}


\begin{table}[h]
\centering
\caption{\textbf{Parameter table:} 
Values of the parameters used for the initial RCD template and obtained from measurements of the fabricated RCD-PyC structures after annealing. 
The geometric surface area ($GSA$) and volume $V$ calculated from the corresponding CAD models are also provided.}
\label{tab:S1_geometry-parameters}
\renewcommand{\arraystretch}{1.3} 
\setlength{\tabcolsep}{12pt} 
\begin{tabular}{|c|c|c|}
\hline
\textbf{Parameter} & \textbf{Template} & \textbf{Post-annealing} \\
\hline
$a$ ($\mu$m)        & 2000          & 630 \\
\hline
$b$ ($\mu$m)        & 1414          & 445 \\
\hline
$d$ ($\mu$m)        & 360           & 160 \\
\hline
$D$ ($\mu$m)        & 17400         & 5500 \\
\hline
$H$ ($\mu$m)        & 6000          & 1260 \\
\hline
$N_{xy}$            & 8.7           & 8.7 \\
\hline
$N_{z}$             & 3             & 2 \\
\hline
$GSA$ ($\mu$m$^{2}$) & $2\times10^{9}$ & $2\times10^{8}$ \\
\hline
$V$ ($\mu$m$^{3}$)   & $2\times10^{11}$ & $8\times10^{9}$ \\
\hline
\end{tabular}
\end{table}

\begin{table}[h]
\caption{\label{tab:S2_ECSA_over_GSA}\textbf{Surface area enhancement factors corresponding to geometric surface area (GSA) and electrochemical active surface area (ECSA).} 
The ECSA values were calculated using C\textsubscript{dl} and C\textsubscript{s} derived from CV in the non-Faradaic region.}
\begin{indented}
\item[]\begin{tblr}{
    hlines,
    vlines,
    cell{2}{1} = {r=3}{},
    cell{5}{1} = {r=3}{},
    width=\linewidth,
    cells={valign=m},
    colspec={Q[c]Q[c]Q[co=1,c]Q[co=1,c]Q[co=1.2,c]},
}
\textbf{Electrolyte} & \textbf{Sample} & \textbf{GSA (cm$^{2}$)} & \textbf{ECSA (cm$^{2}$)} & \textbf{Enhancement factor (ECSA:GSA)} \\
0.5 M \ch{H2SO4} & CN/Ni/RCD-PyC & 2 & 170 & \ECSAoverGSAacidicCN{} \\
& Ni/RCD-PyC    & 2 & 20  & \ECSAoverGSAacidicNi{} \\
& Bare RCD-PyC  & 2 & 2   & \ECSAoverGSAacidicBare{} \\
1 M \ch{KOH} & CN/Ni/RCD-PyC & 2 & 155 & \ECSAoverGSAalkalineCN{} \\
& Ni/RCD-PyC    & 2 & 48  & \ECSAoverGSAalkalineNi{} \\
& Bare RCD-PyC  & 2 & 5   & \ECSAoverGSAalkalineBare{} \\
\end{tblr}
\end{indented}
\end{table}



\clearpage{}



\clearpage{}

\section*{EIS Fit Parameters}
\begin{table}[h]
\centering
\caption{\textbf{EIS fit parameters.}}
\label{tab:S3_EIS_fit_parameters}
\renewcommand{\arraystretch}{1.3}
\setlength{\tabcolsep}{12pt}
\begin{tabular}{|c|c|c|c|c|}
\hline
\textbf{Electrolyte} & \textbf{Sample} & \boldmath$R_s$ & \boldmath$R_{CT}$ & \textbf{CPE} \\
\hline
\multirow{3}{*}{0.5 M \ch{H2SO4}} 
& CN/Ni/RCD-PyC & 7.6  & 12.1  & 0.682 \\
\cline{2-5}
& Ni/RCD-PyC    & 12.4 & 36    & 0.437 \\
\cline{2-5}
& Bare RCD-PyC  & 20.5 & 430   & 0.998 \\
\hline
\multirow{3}{*}{1 M \ch{KOH}} 
& CN/Ni/RCD-PyC & 10.5 & 152   & 0.996 \\
\cline{2-5}
& Ni/RCD-PyC    & 7.83 & 171   & 0.992 \\
\cline{2-5}
& Bare RCD-PyC  & 8.0  & 1810  & 0.688 \\
\hline
\end{tabular}
\end{table}


\begin{figure}[h!]
    \centering
\includegraphics[width=\linewidth]{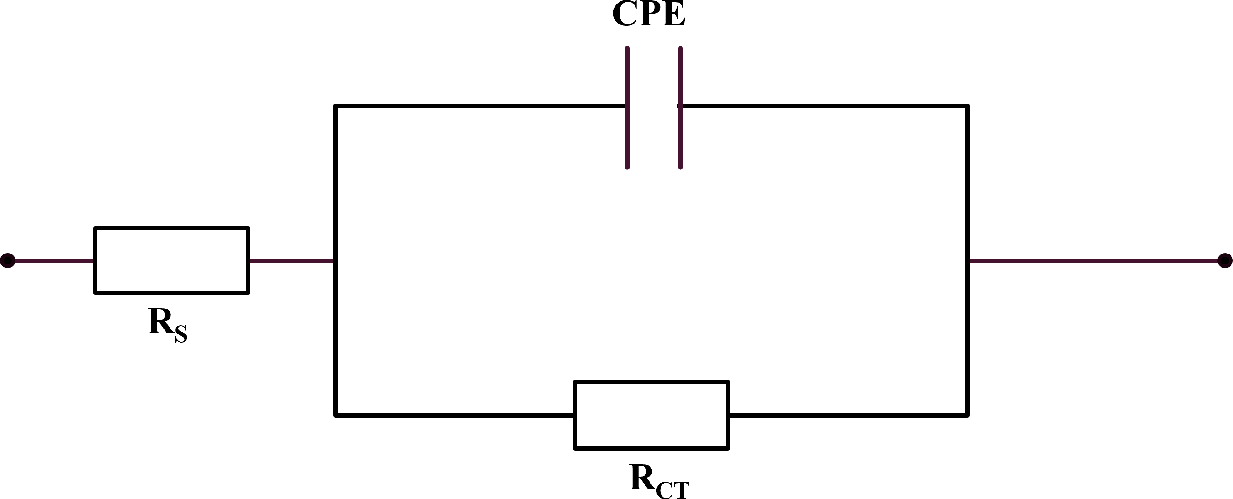}
    \caption{\textbf{Equivalent circuit used for fitting EIS data.} The equivalent circuit consists of three components corresponding to solution resistance ($R_s$), charge transfer resistance ($R_{CT}$) and a constant phase element (CPE).}
    \label{fig:S11_EIS_equivalent_circuit}
\end{figure}

\clearpage
\section*{Cyclic Stability}

\begin{figure}[h!]
    \centering
\includegraphics[width=\linewidth]{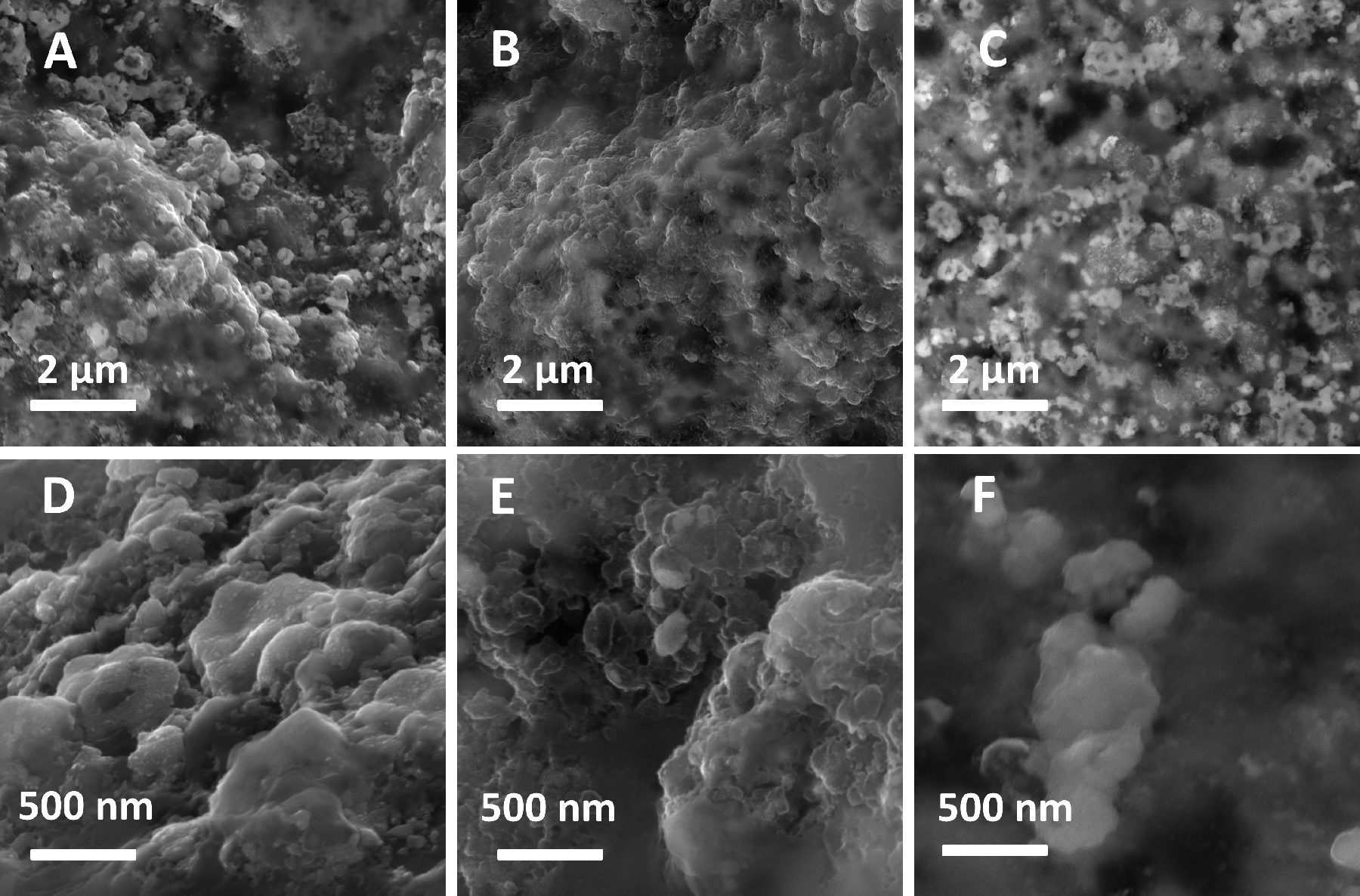}
    \caption{\textbf{Comparison of the CN/Ni/RCD-PyC electrode morphology before and after CV cyclic stability analysis:} A,D: before testing; B,E: after testing in acidic electrolytes for 2000 CV cycles; C,F: after testing in alkaline electrolytes for 2000 CV cycles.}
    \label{fig:S12_SEM_cyclic_stability}
\end{figure}

%
%
%
%
%






\clearpage{}

\printbibliography 